\documentclass[prd, twocolumn, nofootinbib, floatfix]{revtex4-1}

\usepackage[mathscr]{euscript}
\usepackage{amsmath}
\usepackage{graphicx}
\usepackage{dcolumn}
\usepackage{bm}
\usepackage{epsfig}
\usepackage{amssymb,latexsym,mathrsfs}
\usepackage{graphicx}
\usepackage{color}
\usepackage{hyperref}
\usepackage{float}
\usepackage{diagbox}

\definecolor{vividviolet}{rgb}{0.62, 0.0, 1.0}
\definecolor{amaranth}{rgb}{0.9, 0.17, 0.31}
\definecolor{palatinateblue}{rgb}{0.15, 0.23, 0.89}
\definecolor{brightpink}{rgb}{1.0, 0.0, 0.5}
\definecolor{cornflowerblue}{rgb}{0.39, 0.58, 0.93}
\definecolor{deepcarminepink}{rgb}{0.94, 0.19, 0.22}
\definecolor{radicalred}{rgb}{1.0, 0.21, 0.37}

\hypersetup{ linktoc=all,
    colorlinks, linkcolor={palatinateblue},
    citecolor={brightpink}, urlcolor={amaranth}
}
\newcommand{\changeurlcolor}[1]{\hypersetup{urlcolor=#1}}
\newcommand{\be}{\begin{equation}}
\newcommand{\ee}{\end{equation}}
\newcommand{\bs}{\begin{split}} 
\newcommand{\bea}{\begin{eqnarray}}
\newcommand{\eea}{\end{eqnarray}}

\renewcommand{\d}[1]{\ensuremath{\operatorname{d}\!{#1}}}

\begin{document}

\title{Particle Spectrum of the Reissner-Nordstr\"om Black Hole}
\author{Michael R.R. Good${}^{1}$}
\email{michael.good@nu.edu.kz}
\author{Yen Chin Ong${}^{2}$}
\email{ycong@yzu.edu.cn}
\affiliation{${}^1$Physics Department \& Energetic Cosmos Laboratory, \\
Nazarbayev University, Nur-Sultan, Kazakhstan\\
${}^2$Center for Gravitation and Cosmology, College of Physical Science and Technology,\\ Yangzhou University, Yangzhou, China\\
}

\begin{abstract} 
The Reissner-Nordstr\"om black hole - moving mirror correspondence is solved.  The beta coefficients reveal that charge makes a black hole radiate fewer particles (neutral massless scalars) per frequency.  An old Reissner-Nordstr\"om black hole emits particles in an explicit Planck distribution with temperature corresponding to the surface gravity of its outer horizon.
\end{abstract} 

\date{\today} 

\maketitle
\section{Introduction} 
Does the Reissner-Nordstr\"om (RN) black hole radiate fewer neutral massless particles than the Schwarzschild black hole?  
It is well-known that charged black holes are colder and smaller than their neutral cousins (of the same $M$).  
Consider the range of sizes of the outer horizon of a RN black hole and the
temperature range,
\be \frac{1}{2} < \frac{r_+}{r_s} < 1, \quad 0 < \frac{T_+}{T_s} < 1.\ee
where the $s$-subscript refers to Schwarzschild, $r_s \equiv 2M$ and $T_s = (4\pi r_s)^{-1}$.  The inner and outer horizons of the RN black holes are
\be r_{\pm} = \frac{1}{2}\left(r_s \pm \sqrt{r_s^2 -4 r_q^2}\right),\ee
which are dependent on the two parameters $(M,Q)$, where $r_q^2\equiv Q^2$, and the temperature of the outer horizon is \cite{springy}
\be  T_+ = \frac{\kappa_+}{2\pi} = \frac{2r_+-r_s}{4\pi r_+^2} = T_s - T_s\left(\frac{\bar{T}_+}{T_s}-1\right)^2,\ee
where $\bar{T}_+ := (4\pi r_+)^{-1} > T_s$. 

In this note, we consider whether the number of neutral massless particles radiated is reduced too.  
By modeling the RN black hole by a moving mirror \cite{DeWitt, Davies1,Davies2}, we solve for the beta coefficients for all times, which is the spectrum of the collapsing star \cite{RNprob} up to gray body factors and different dimensionality (for a recent review see \cite{Dodonov:2020eto}).  

This paper is organized as follows:  Section \ref{sec:RN} contains the RN metric and matching condition for collapse.  Section \ref{sec:Traj} reveals the moving mirror, and computes the dynamics resulting in asymptotic infinite acceleration.  Section \ref{sec:particles} demonstrates the all-time spectrum and compares with the Schwarzschild and extremal Reissner-Nordstr\"om spectra.  In Section \ref{sec:conc}, we conclude.  Appendix \ref{appx} has an elementary model with reduced particle creation.   Units are $G = \hbar = c  = 1$.

\section{Reissner-Nordstr\"om Metric} \label{sec:RN}
The outside metric of the RN collapse system, is given by the RN geometry,
\be \d s^2 = -f(r)\d t^2 + f(r)^{-1}\d r^2 + r^2\d\Omega^2,\ee
where our $f(r) \equiv f_q$ is given by
\be f_q = 1-\frac{r_s}{r} + \frac{r_q^2}{r^2} = \frac{(r-r_+)(r-r_-)}{r^2}.\label{metric} \ee
For a double null coordinate system $(u,v)$, utilizing $u = t-r^*$, and $v = t+r^*$, with the appropriate tortoise coordinate \cite{Fabbri}, 
\be r^* = r+ \frac{1}{2\kappa_+}\ln \left|\frac{r-r_{+}}{r_{+}}\right|+\frac{1}{2\kappa_-}\ln \left|\frac{r-r_{-}}{r_{-}}\right|,\ee
where the surface gravities are defined as, 
\be \kappa_\pm \equiv \frac{r_\pm - r_\mp}{2r_\pm^2},\ee
one has the metric for the outside collapse geometry,
\be \d s^2 =-f_q\; \d u \d v + r^2 \d\Omega^2.\ee
The matching condition (see e.g. \cite{purity,Fabbri}) between the flat interior geometry, described by the interior coordinates, $U=T-r$, and $V=T+r$, is the trajectory of the incipient black hole origin, expressed in terms of the exterior function, $u(U)$, dependent on interior coordinate, $U$:
\be u(U) = U-\frac{\ln \left|-\frac{2 r_++U-v_0}{2 r_+}\right|}{\kappa _+}-\frac{\ln \left|-\frac{2 r_-+U-v_0}{2 r_-}\right|}{\kappa _-}.\label{matchunset}\ee
This matching, $r^*(r=(v_0 -U)/2) = (v_0-u)/2$, happens along the shell, $v_0$.  Here $v_0 -v_H\equiv 2r_+$ because $u\rightarrow +\infty$ at $U=v_H$.  We set the horizon to zero, $v_H=0$:
\be u(U) = U - \frac{1}{\kappa_+}\ln \left|-\frac{U}{2r_{+}}\right|-\frac{1}{\kappa_-}\ln \left|\frac{r_+}{r_-}-1-\frac{U}{2r_-}\right|.\label{match}\ee
Regularity of the field at $r=0$ acts like a moving mirror (no field behind $r<0$), revealing the form of field modes, such that $U\leftrightarrow v$ identification can be made for the Doppler-shifted right movers.  The mirror trajectory, $f(v) \leftrightarrow u(U)$ is then a known function of advanced time, which we examine in the next section.

\section{Trajectory and Dynamics}\label{sec:Traj}
\begin{figure}[ht]
\centering 
\includegraphics[width=3.4in]{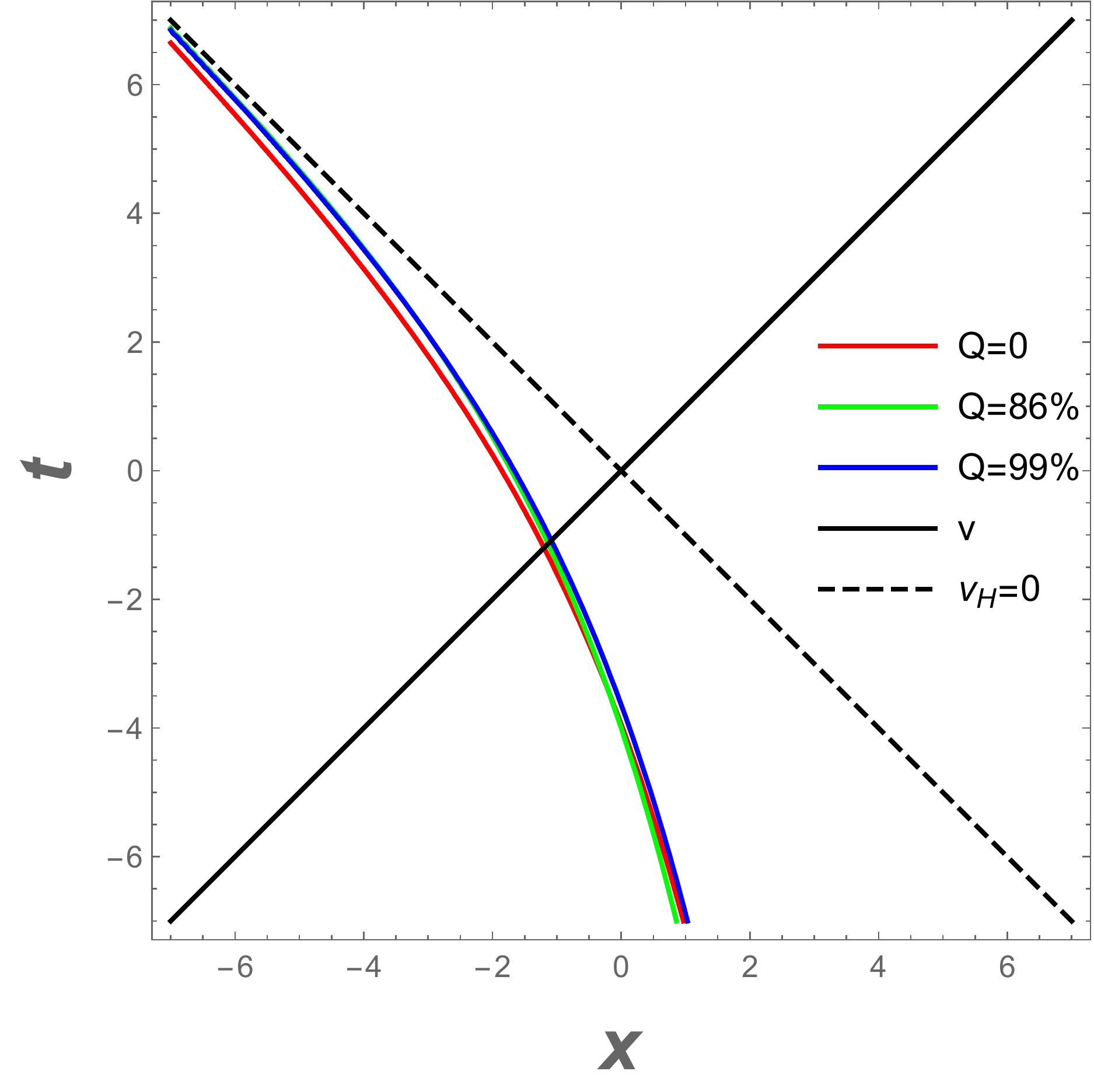} 
\caption{Trajectories Eq.~(\ref{f(v)general}), in a spacetime plot.  The horizon has been set to $v_H = 0$, the charge is  $Q=0$, $Q=\sqrt{3}/2=86\%$ and $Q=99\%$, for $M=1$.  The mirrors starts asymptotically static, but have infinite proper acceleration in the far future. As can be seen, effectively, charge does not significantly alter the qualitative behavior of the curves. 
}
\label{fig:SpacetimePlot} 
\end{figure}

\begin{figure}[ht]
\centering 
\includegraphics[width=3.2in]{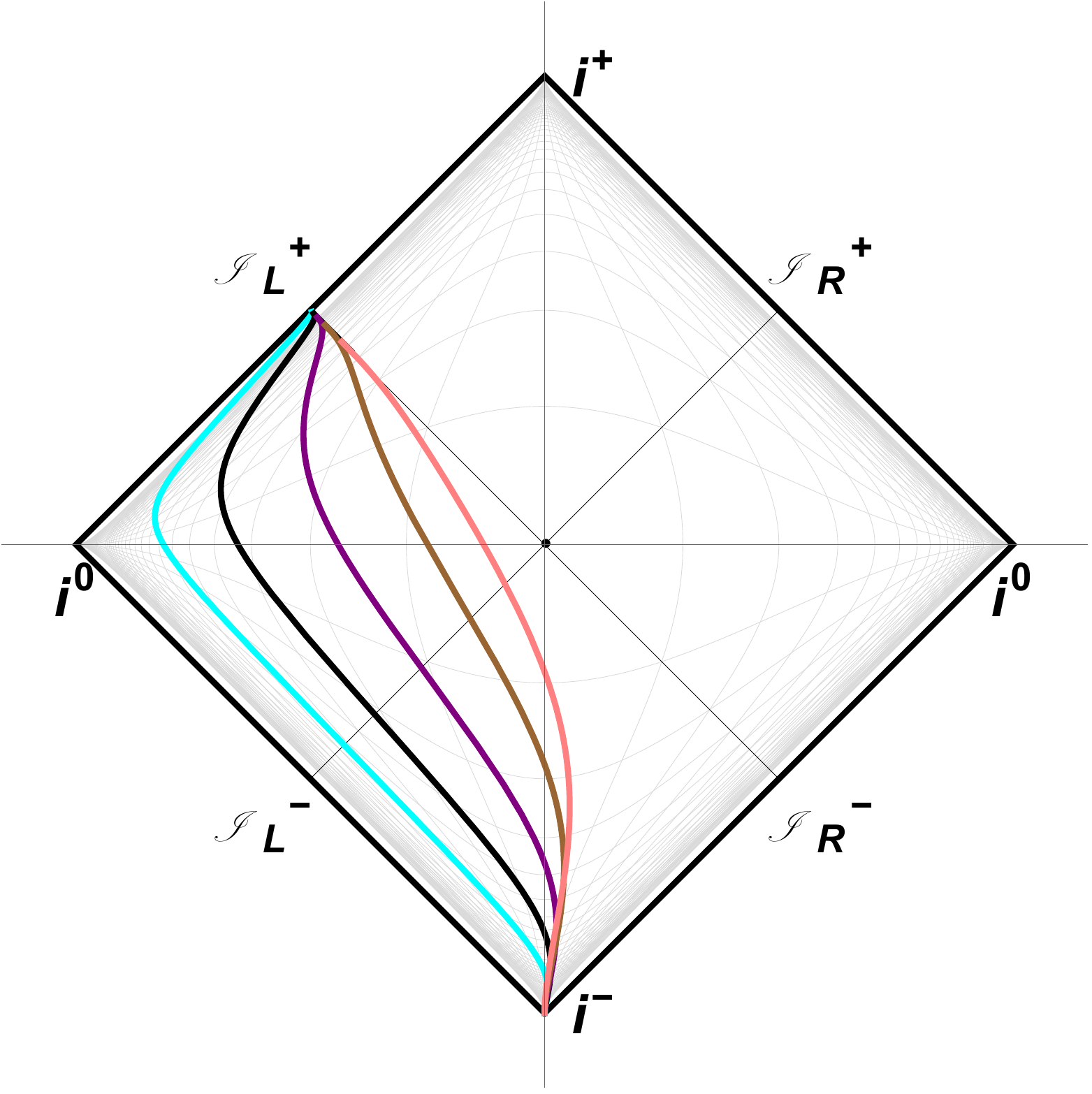} 
\caption{The trajectory Eq.~(\ref{f(v)general}) in a Penrose conformal diagram.  The horizon has been set to $v_H = 0$. For illustrative emphasis on the asymptotic behavior, the ratio $Q/M = 86\%$ is fixed, but the mass is changing, $M=2, 1, 2^{-1}, 4^{-1}, 8^{-1}$; for cyan, black, purple, brown, pink, respectively.  The mass dramatically affects the conformal trajectory, demonstrating the horizon approach is more rapid at late times for higher value of the mass. 
}
\label{fig:PenrosePlot} 
\end{figure}

Let us study massless scalar field in $(1+1)$-dimensional Minkowski spacetime (following  \cite{signatures}). The corresponding RN moving mirror trajectory from Eq.~(\ref{match}) is 
\be f(v) =v-\frac{\ln \left(\frac{1}{2} \sqrt{\frac{v^2}{r_+^2}}\right)}{\kappa _+}-\frac{\ln \left(\frac{1}{2} \sqrt{\frac{\left(2 r_--2 r_++v\right){}^2}{r_-^2}}\right)}{\kappa _-},\label{f(v)general}\ee
expressed in null coordinates $(u,v)$ where $f(v)$ is a function of null coordinate advanced time $v$. This is a rather involved dynamics even though the system only has two parameters: $(M,Q)$. Setting $M=1$, but more notably, setting the charge to a specific value that is particularly convenient:  $Q = \sqrt{3}/2 = 86\%$, simplifies the dynamics for illustration ($r_- = 1/2$, $r_+=3/2$):
\be f(v) =v -\frac{9}{4} \ln \left(\frac{v^2}{9}\right)+\frac{1}{4} \ln (v-2)^2.\label{f(v)}\ee
Keep in mind the horizon has been set to $v_H=0$, and so $v$ spans $-\infty < v < v_H$.  The trajectory in spacetime coordinates is plotted in a spacetime plot Fig.~(\ref{fig:SpacetimePlot}).  A conformal diagram of the accelerated boundary is given in Fig.~(\ref{fig:PenrosePlot}).  

The rapidity, $\eta = -\frac{1}{2}\ln f'(v)$; and proper acceleration, $\alpha = e^{\eta(v)}\eta'(v)$; are monotonic functions that diverge in the limit that advanced time approaches the horizon time, $v\rightarrow v_H=0$.  The limit in the far past, $v\rightarrow -\infty$ is $(\eta,\alpha) \rightarrow 0$;  the mirror is past asymptotically static.  The mirror rapidly travels left, off to the speed of light. 

In the limit that charge goes to zero, $Q\rightarrow 0$, the function, $u(U)\leftrightarrow f(v)$, of Eq.~(\ref{match}), recovers the Schwarzshild mirror \cite{Good:2016MRB,Good:2016LECOSPA,MG14one,MG14two}, as one expects:
\be f(v) = v- 4M \ln\left|\frac{v_H-v}{4M}\right|.\ee
We will utilize the Schwarzschild surface gravity, $\kappa \equiv (4M)^{-1}$ in the following.

\section{Spectrum and Particles}\label{sec:particles}
The beta Bogoliubov coefficient can be found via \cite{horizonless},
\be \beta_{\omega\omega'} = -n_0\int_{-\infty}^{v_H} \d v ~e^{-i \omega' v -i \omega f(v)}\left(\omega f'(v)-\omega'\right),\label{betaint}\ee
by setting the horizon $v_H=0$ for convenience and definiteness, (horizon position will not affect the spectrum because of complex conjugation). The normalization factor is $n_0^{-1}= 4\pi\sqrt{\omega\omega'}$. We define

\be a \equiv \frac{i\omega}{\kappa_+},\quad b\equiv \frac{i\omega}{\kappa},\quad c \equiv -\frac{i\omega_p}{\bar{\kappa}},\quad d\equiv -\frac{i\omega}{\kappa_-},\ee
in which we have the RN and Schwarzschild surface gravities, $(\kappa_\pm,\kappa)$ and also, $ \bar{\kappa} \equiv (2(r_+ - r_-))^{-1}$. 
The result for the beta coefficient is the lengthy expression:
\be \beta_{\omega\omega'}= \frac{\sqrt{\omega '}}{2  \omega _p \sqrt{\omega }} \text{csch}\left(\frac{ \pi  \omega }{\kappa}\right)\bar{A}, \label{ans} \ee
where $\omega_p \equiv \omega + \omega'$, and
\be \bar{A} \equiv x(A-bB)+y(C-c D) ,  \ee
in which
\be x \equiv -\frac{\Gamma \left(a\right)}{\Gamma \left(d\right)}\left(\frac{r_-}{r_+}\right)^{d} \left(\frac{1}{2\bar{\kappa }r_+}\right)^{b}, \ee
\be y \equiv \left(\frac{r_-}{r_+}\right)^a \left(\frac{1}{-2ir_- \omega_p}\right)^b,\ee
while
\be A \equiv \, _1\tilde{F}_1\left(a;b;c\right),\quad B \equiv \, _1\tilde{F}_1\left(a;1+b;c\right),\ee
\be C \equiv \, _1\tilde{F}_1\left(1+d;-b;c\right),\; D \equiv \, _1\tilde{F}_1\left(1+d;1-b;c\right),\ee
are the regularized confluent hypergeometric functions.  

\begin{figure}[ht]
\centering 
\includegraphics[width=3.2in]{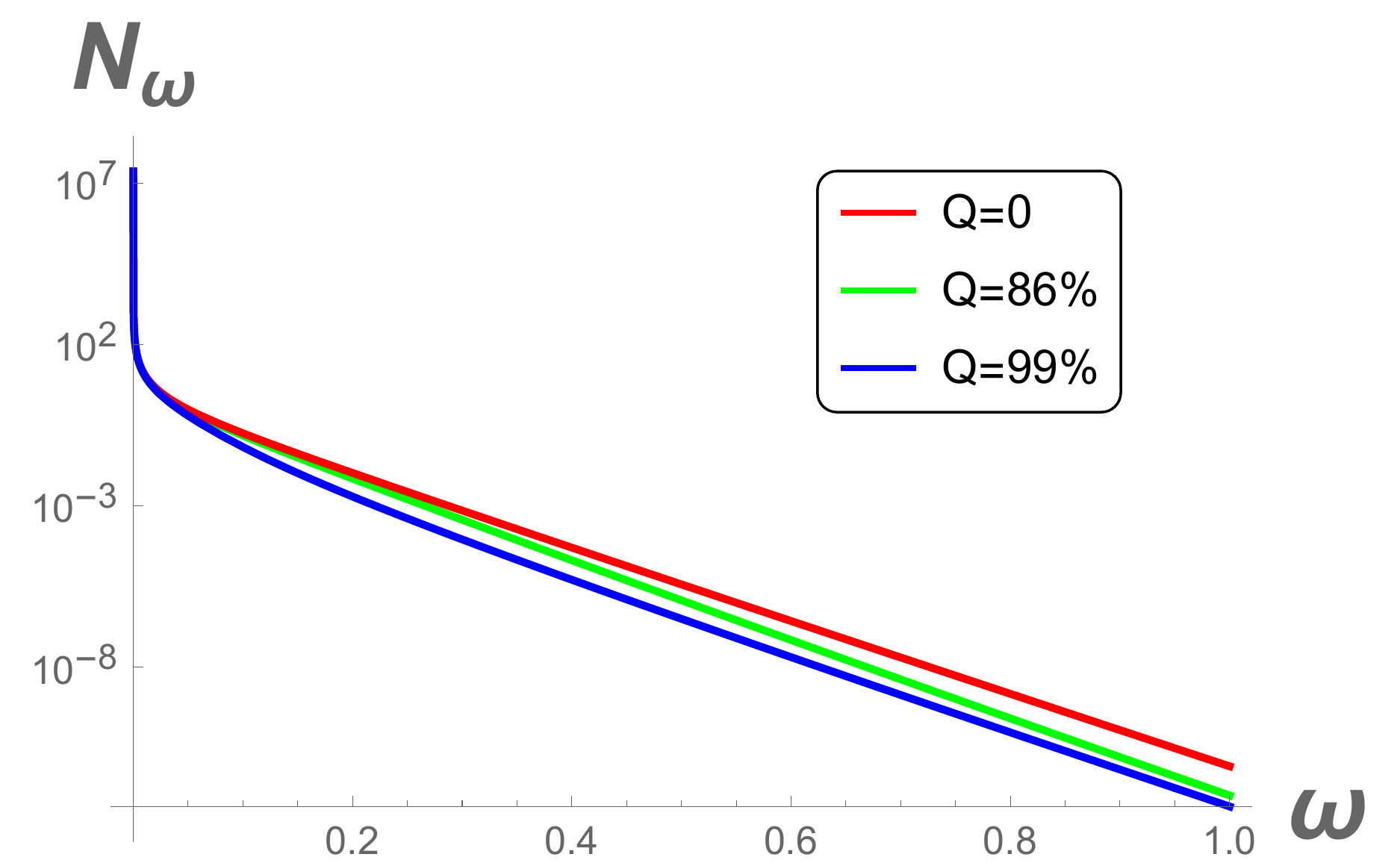} \caption{ The particle spectrum, Eq.~(\ref{N(w)}), in a semi-log plot, demonstrating less particles emitted as the charge is increased.  An infrared divergence, for $\omega\rightarrow 0$ is present, which signals the usual infinite total particle count due to the soft particles at $\omega =0$.  The plot includes the Schwarzschild spectrum, $Q=0$, Eq.~(\ref{Schw}).  Here we set $M=1$.       
}
\label{fig:ParticleSpecPlot} 
\end{figure}

To obtain the spectrum, we complex conjugate,
\be N_{\omega \omega'} \equiv |\beta_{\omega\omega'}^{\textrm{RN}}|^2. \ee
which is the particle count per mode squared.  The spectrum, which is the main result of this paper, is then,
\be N_\omega = \int_0^\infty  N_{\omega \omega'} d\omega',\label{N(w)}\ee
which is plotted in Fig.~(\ref{fig:ParticleSpecPlot}).

\subsection*{Spectrum in Schwarzschild Limit}
The spectrum for the beta coefficient squared for the RN mirror in the limit that the charge, $Q\rightarrow 0$, is the limit where the inner radius goes to zero, $r_- \rightarrow 0$.  The result is confirmed to be
\be \lim_{r_-\rightarrow 0}|\beta_{\omega\omega'}^{\textrm{RN}}|^2 =|\beta_{\omega\omega'}^{\textrm{S}}|^2= \frac{r_s \omega '}{\pi  \left(e^{4 \pi  r_s \omega }-1\right) \left(\omega '+\omega \right)^2}.\label{Schw}\ee
In the high frequency regime, where the modes are extremely red-shifted, $\omega'\gg \omega$, one has $N_{\omega \omega'} :=|\beta_{\omega\omega'}|^2$,
\be N_{\omega \omega'} = \frac{1 }{2\pi\kappa \omega'}\frac{1}{e^{\omega/T_s }-1}.\ee
This confirms that the particle spectrum of Eq.~(\ref{ans}), gives the known answer \cite{CW} in the zero charge limit.\\    

\subsection*{Spectrum in the Extremal Limit}
In the opposite limit for high charge,
the results should conform to the extremal Reissner-Nordstr\"om (ERN) black hole spectrum \cite{Good:2020nmz},
\be |\beta_{\omega\omega'}|^2 = \frac{4 M^2  e^{-4 M \pi  \omega } \omega ' }{\pi ^2 \omega _p}\left|K_a\left(4 M \sqrt{\omega  \omega _p}\right)\right|^2\label{ERNbeta2},\ee
in the correct limit, where $a \equiv 1+4 M i\omega$, and $K_a$ is the modified Bessel function of the second kind. Indeed, taking the maximal charge limit $Q^2\rightarrow M^2$ or $r_\pm \rightarrow r_s/2 = M$ of the integrand of Eq.~(\ref{betaint}) and then integrating over advanced time, the ERN spectrum is obtained.  In the high frequency limit, this amounts to 
\be |\beta_{\omega\omega'}|^2 = \frac{4M^2}{\pi^2} \left| K_1\left(4M \sqrt{\omega  \omega '}\right)\right|^2,\label{UAbeta2}\ee
which, for a uniformly accelerated mirror \cite{pagefulling}, is the spectrum with $\kappa_\textrm{UA} \equiv 1/(2M)$, distinctly non-thermal \cite{Davies1,Davies2,B&D}, as expected since the ERN black hole has ``zero'' (i.e. undefined) Hawking temperature \cite{Anderson:2000pg, LRS}.

\subsection*{Spectrum in the Late-Time Limit}
Finally, one can check that at late time, thermal behavior is obtained for the beta coefficient squared, as expected:
\be \lim_{\omega'\gg\omega} |\beta_{\omega\omega'}^{\textrm{RN}}|^2 = \frac{1}{2\pi\kappa_+ \omega' } \frac{1}{e^{\omega/T_+ }-1 },\ee
with temperature, $T_+ = \kappa_+/(2\pi)$.

\section{Conclusions}\label{sec:conc}

We have solved for the particle spectrum of the RN black hole by use of the moving mirror model solving for the beta Bogoliubov coefficients.  In the limits of zero charge and maximum charge, the Schwarzshild and the ERN results are obtained, respectively. In the high frequency regime corresponding to late time the spectrum is thermal with temperature $T_+ = \kappa_+/(2\pi)$.  The result demonstrates that during the formation of the RN black hole, charge inhibits the radiation of massless scalar neutral particles relative to an equal mass Schwarzschild black hole.  The thermal count establishes the supposition (and confirms explicitly through particle production) that the distribution of particles from a RN black hole is the Planck spectrum.  
     

\appendix

\section{Truncated Method}\label{appx}
The essential physics is actually encapsulated in the truncated trajectory:
\be f(v)_{\textrm{Trun}} = v - \frac{1}{\kappa_+}\ln \left|\frac{v}{2r_+}\right|,\ee
which results in a significant simplification. This motion is a good preliminary model for investigation of a spectrum that has the correct Schwarzschild limit and late-time limit. The corresponding exterior metric is 
\be \d s^2_{\textrm{Trun}} = \left(1-\frac{1}{2\kappa_+ r}\right)\d u\d v + r^2\d\Omega^2,\ee
which has surface gravity $\kappa_+$ like the RN metric.  But unlike the RN metric this is a vacuum solution, $G_{\mu\nu}=0$ (so one could argue that $Q$ is no longer the electric charge), with horizon at $r=1/(2\kappa_+) = r_+^2/(r_+-r_-) \neq r_+$. Eq.~(\ref{betaint}) results in a spectrum that is simply,
\be |\beta_{\omega\omega'}^{\textrm{Trun}}|^2 = \frac{ \omega '}{2\pi \kappa_+  \left(\omega '+\omega \right)^2} \frac{1}{e^{\omega/T_+ }-1 }.\label{Trun}\ee
The Schwarzschild zero charge limit holds,
\be \lim_{Q\rightarrow 0} |\beta_{\omega\omega'}^{\textrm{Trun}}|^2 = |\beta_{\omega\omega'}^{\textrm{S}}|^2,\ee
and the late-time thermal limit also holds,
\be \lim_{\omega'\gg\omega} |\beta_{\omega\omega'}^{\textrm{Trun}}|^2 = \frac{1}{2\pi\kappa_+ \omega' } \frac{1}{e^{\omega/T_+ }-1 },\ee
with temperature, 
\be T_+ = \frac{\kappa_+}{2\pi}.\ee
Since $\kappa_+<\kappa$, the particle production is mitigated by the presence of charge. \newline

\acknowledgments 

Much thanks to Paul Anderson for input on an early draft of this manuscript.  Funding from state-targeted program ``Center of Excellence for Fundamental and Applied Physics" (BR05236454) by the Ministry of Education and Science of the Republic of Kazakhstan is acknowledged. MG is also funded by the ORAU FY2018-SGP-1-STMM Faculty Development Competitive Research Grant No. 090118FD5350 at Nazarbayev University. YCO thanks the National Natural Science Foundation of China (No.11922508, No.11705162) and the Natural Science Foundation of Jiangsu Province (No.BK20170479) for funding support.




\end{document}